\documentclass[aps,prl,twocolumn,showpacs]{revtex4}
\usepackage{graphicx}

\begin{document}

\title{Conductance of Pd-H nanojunctions}

\author{Sz.~Csonka, A.~Halbritter and G.~Mih\'aly}
\affiliation{Electron Transport Research Group of the Hungarian
Academy of Science and Department of Physics, Budapest University
of Technology and Economics, 1111 Budapest, Hungary}

\author{O.I.~Shklyarevskii, S.~Speller, and H.~van~Kempen}
\affiliation{NSRIM, University of Nijmegen, Toernooiveld 1, 6525
ED Nijmegen, the Netherlands}

\date{\today}

\begin{abstract}
Results of an experimental study of palladium nanojunctions in
hydrogen environment are presented. Two new hydrogen-related
atomic configurations are found, which have a conductances of
$\sim 0.5$ and $\sim 1$ quantum unit ($2e^2/h$). Phonon spectrum
measurements demonstrate that these configurations are situated
between electrodes containing dissolved hydrogen. The crucial
differences compared to the previously studied Pt-H$_2$ junctions,
and the possible microscopic realizations of the new
configurations in palladium-hydrogen atomic-sized contacts are
discussed.
\end{abstract}


\maketitle

In the recent years a great progress has been achieved in
understanding the conduction properties of matter at atomic scale.
A large variety of experiments performed by Scanning Tunnelling
Microscopes (STM) and the Mechanically Controllable Break Junction
(MCBJ) technique have provided a comprehensive picture about the
nature of conductance in monoatomic metallic junctions
\cite{Agrait2003}. Even more recently it was shown, that these
techniques can also be used to investigate the conductance through
{\it molecules} between metallic electrodes
\cite{Reed1997,Reichert2002}. Such measurements, however, are
strongly influenced by the interaction between the molecule and
the electrodes, and this interaction is poorly controllable in the
experiment. Furthermore, it is difficult to make certain that a
single molecule forms the junction.

Promising results have been achieved by studying junctions with
the simplest molecule, H$_2$. It was shown by Smit et al.\
\cite{Smit2002} that a single hydrogen molecule can form a stable
bridge between platinum electrodes, which has a conductance of one
quantum unit ($G_0=2e^2/h$), carried by a single channel. This
configuration is naturally created during the rupture of a
platinum nanojunction placed in H$_2$ environment. The conductance
of this molecular bridge is not an intrinsic property of H$_2$,
but it results from the fine details of the connection to the
electrodes. Therefore, both the atomic arrangement created by the
hydrogen and its conductance is expected to be material dependent.
For instance in gold junctions a new atomic configuration with a
conductance of $\sim 0.5 \, G_0$ arises, which has been related to
an atomic Au wire distorted due to the adsorption of hydrogen
\cite{Csonka2003}.

 The interaction of hydrogen with the atomic-sized contact
is an especially interesting problem for palladium junctions, as
Pd is highly reactive with respect to H$_2$. The study of Smit et
al.\ on the Pt-H$_2$ system \cite{Smit2002} noted that Pd behaves
similarly in hydrogen surrounding as the isoelectronic platinum.
Our present work, however, shows that the palladium-hydrogen
system has a more complex behavior, which is attributed to the
dissolution of hydrogen in the electrodes.

\begin{figure}
\includegraphics[width=0.45\columnwidth]{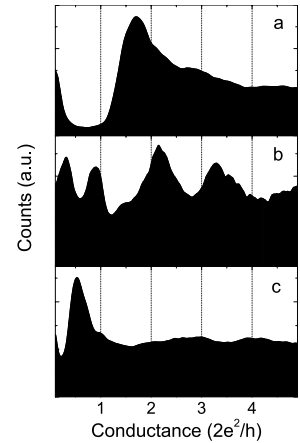}
\caption{\it Conductance histograms for clean Pd junction (a) and
Pd in hydrogen atmosphere (b), (c) measured at $V=150$\,mV and
$T=5$\,K.} \label{histograms.fig}
\end{figure}

We have performed measurements on high purity Pd samples. The
contacts were created by the MCBJ technique in cryogenic
environment. The junctions were studied by conductance histograms
\cite{Agrait2003}, which are constructed from conductance vs.\
electrode separation traces recorded during thousands of ruptures.
The peaks in the histogram reflect the stable atomic
configurations of the junction.

The conductance histogram of Pd measured in high vacuum is
presented in Fig.~\ref{histograms.fig}(a). It shows a well defined
peak at $G\simeq 1.8\,G_0$ corresponding to the conductance of a
monoatomic Pd contact, which can have up to 5 partially open
conductance channels \cite{Cuevas1998}. The shape of the histogram
is insensitive to the experimental parameters (bias voltage and
temperature) in the range of the measurements ($V=10-300$\,mV,
$T=4.2-30$\,K).

A remarkable change is observed in the histogram if a small amount
of H$_2$ is admitted to the vacuum pot \cite{note1}. Due to the
presence of hydrogen the peak characteristic to a single-atom Pd
contact disappears, and two new peaks appear in the histogram at
$\sim 0.5\,G_0$ and at $\sim 1\,G_0$, respectively
[Fig.~\ref{histograms.fig}(b)]. We note that the first peak has a
rather indefinite position in the region $\sim 0.3-0.6\,G_0$. The
second peak at $\sim 0.9-1\,G_0$ may disappear if the amount of
hydrogen is increased, as shown in Fig.~\ref{histograms.fig}(c).

These histograms demonstrate that two new stable atomic
configurations arise due to the adsorption of hydrogen on
palladium contacts. The relation of these two configurations to
each other can be studied by statistical analysis following the
method introduced in our previous work \cite{Csonka2003}. We found
that these two configurations are uncorrelated, which means that
the appearance  of the first or the second configuration during
the same rupture are independent events. It shows that both
configurations arise between the same electrodes, but the
appearance of the configuration at $\sim 1\,G_0$ neither helps nor
hampers the appearance of the configuration at $\sim 0.5\,G_0$
during further retraction.

\begin{figure}[b!]
\includegraphics[width=0.7\columnwidth]{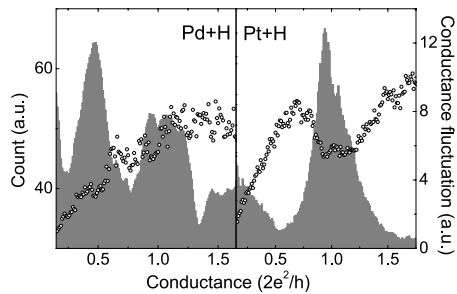}
\caption{\it The left panel shows the conductance fluctuation
measurement on Pd-H junctions exhibiting both new configurations.
As a reference the conductance fluctuation in Pt-H contacts is
shown in the right panel. In the background the conductance
histograms are presented for the same data sets.}
\label{condfluct.fig}
\end{figure}

Additional information about the conductance channels of the
configuration at $\sim 1\,G_0$ can be obtained by conductance
fluctuation measurements. In Pt junctions the suppression of
quantum interference at $1\,G_0$ gave evidence that the
hydrogen-related configuration has a single conductance channel
\cite{Smit2002}. As a sharp contrast we have found that in Pd
junctions the conductance fluctuations are not suppressed (see
Fig.~\ref{condfluct.fig}), thus the configuration with $G=1\,G_0$
has more than one open channels.

We have studied the structure of the electrodes around the new
atomic-sized configurations as well. After pressing the electrodes
together to a mesoscopic size ($\approx 100\,G_0$) the vibration
modes of the junction were determined from the point contact (PC)
spectrum \cite{Jansen1980}, i.e. the second derivative of the
$I-V$ curve recorded by standard lock-in technique.

\begin{figure}
\includegraphics[width=0.8\columnwidth]{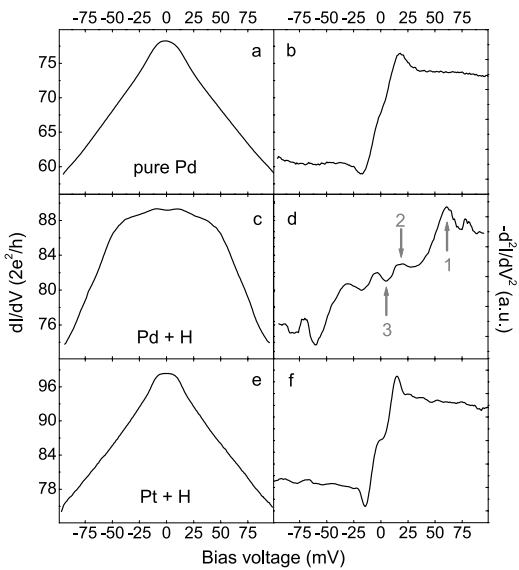}
\caption{\it The left/right panels show the first/second
derivative of the $I(V)$ curves for mesoscopic junctions of pure
Pd (a)/(b), Pd in hydrogen atmosphere (c)/(d) and Pt in hydrogen
atmosphere (e)/(f). In Pt the the spectrum is uneffected by
hydrogen, similar curve is obtained for a pure Pt contact. The
derivatives were measured simultaneously with lock-in technique
using a modulation of $2$\,mV at $T=5$\,K.}
\label{phononspectra.fig}
\end{figure}

The PC spectrum for pure Pd [Fig.~\ref{phononspectra.fig}(b)]
shows a single spectroscopic peak at $V\simeq 15-20$\,mV, which
corresponds to the phonon modes of the Pd crystal in agreement
with earlier results \cite{Caro1981}. In contrast, in the presence
of hydrogen (when the histograms of Fig.~\ref{histograms.fig}(b)
or (c) are observed) the PC spectrum takes a significantly
different form [Fig.~\ref{phononspectra.fig}(d)]. In this case the
spectrum is dominated by a wide peak at $\sim 60$\,mV (arrow 1)
along with a peak corresponding to the phonon modes of the pure Pd
crystal (arrow 2). As a third feature, a zero bias anomaly is also
observed (arrow 3).

These results can be compared with the observations on Pt
contacts, where the vibrational spectrum of a H$_2$ molecular
bridge (Fig.\,2 in Ref.\onlinecite{Smit2002}) shows a closely
similar structure to the PC spectrum in
Fig.~\ref{phononspectra.fig}(d). To this end, we have performed
measurements on mesoscopic-sized Pt junctions as well (see
Fig.~\ref{phononspectra.fig}(f)), which were insensitive to
hydrogen. These measurements show that in Pd contacts the presence
of hydrogen is not only reflected by the new conductance values of
atomic-sized junctions, but it is also markedly shown by the bulk
properties of the electrodes. In contrast, in Pt contacts the
hydrogen acts only on the surface.

Both Pd and Pt are well-known for the chemical adsorption of
hydrogen on the surface. In palladium contacts, however, the
hydrogen can also be dissolved in the bulk crystal due to the
larger space at the octahedral intersticial sites in the fcc
lattice. Indeed, the spectroscopic peak at $\sim 60$\,mV in Pd
junctions coincides with the vibrational modes of dissolved H
atoms in Pd host, in accordance with theoretical calculations
\cite{Rahman1976}, neutron scattering experiments \cite{Rowe1974},
and PC spectroscopy measurements \cite{Caro1983} on
palladium-hydride system (for an overview see
Ref.~\onlinecite{Alefeld1978}). The zero bias anomaly in the PC
spectrum could also be explained by the dissolved hydrogen
\cite{Cox1998}.

Based on the above PC spectroscopy measurements one can conclude
that the new hydrogen-related configurations in Pd junctions are
emerging between electrodes containing dissolved hydrogen. On the
other hand, it should be emphasized that under our experimental
conditions the dissolution of hydrogen is not expected, since the
hydrogen is admitted to the contact in situ at low temperature
($T<20$\,K), where its diffusion in Pd is completely frozen out
\cite{Alefeld1978}. For comparison, we have also performed
measurements on samples that were cooled down after intentional
dissolution of H at room temperature. These experiments provide
the same PC spectrum as those where the hydrogen is added at
cryogenic temperatures. We attribute the unexpected dissolution of
H at low temperature to the local overheating of the junction by
the bias voltage. In the voltage range of the measurements ($\sim
150$\,mV) the contact is easily heated up to $T\sim100$\,K
\cite{Halbritter2002}, where the hydrogen already has a reasonable
diffusion rate ($\sim 1$\,nm/s) \cite{Alefeld1978}. The repeated
ruptures and compressions of the contact during the acquisition of
the histogram may also assist the dissolution of hydrogen from the
contact surface.

The amount of H dissolved in the contact is regulated by the
balance of two parameters: the mobility of hydrogen in the Pd host
and the partial pressure of H$_2$ gas near the junction. The
mobility is controlled by the local overheating whereas the H$_2$
pressure is determined by the amount of hydrogen in the sample
space and the bath temperature. At $T=4.2$\,K the hydrogen is
frozen, thus its partial pressure  is very small
($10^{-6}$\,mBar). The amount of hydrogen gas can be increased
significantly by elevating the temperature to $20$\,K (the boiling
point of hydrogen). If a high bias ($>200$\,mV) is combined with
low bath temperature ($4.2$\,K) then the cryogenic vacuum pumps
out the mobile hydrogen from the junction. In this way the
situation with adsorbed hydrogen (histograms (b) and (c) in
Fig.~\ref{histograms.fig}) can be turned into the case of pure
palladium (histogram (a) in Fig.\ref{histograms.fig}). After this
the electrodes can be ``refilled'' with hydrogen by increasing the
temperature to $20$\,K. As the hydrogen is admitted to the
junction an intermediate state can also be observed for a few
minutes. In this case the presence of hydrogen at the contact is
already reflected by the growth of a single peak in the histogram
at $G=1\,G_0$, however the hydrogen is not yet dissolved in the
electrodes, which is clearly visible in the PC spectroscopy
measurements on mesoscopic junctions.

Summarizing the experimental results, in palladium junctions two
new stable atomic configurations arise due to the adsorption of
hydrogen at $G\sim 0.5 G_0$ and $G\sim 1 G_0$. These two
configurations can independently appear during the same rupture,
between the same electrodes. According to the phonon spectrum
measurements both configurations are situated between electrodes
with dissolved hydrogen. With the aid of previous theoretical
calculations \cite{Heurich2003, Garcia2004} for the
platinum-hydrogen system and the comprehensive knowledge of the
electron structure of bulk palladium hydride \cite{Chan1983}
reasonable microscopic pictures can be suggested for these
configurations, which is discussed below.

\begin{figure}
\centering
\includegraphics[width=\columnwidth]{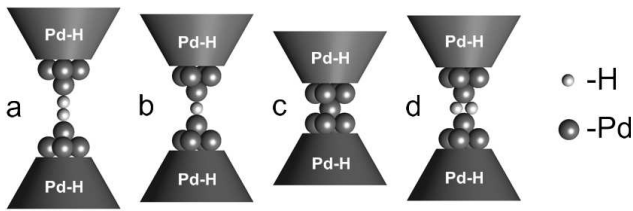}
\caption{\it Possible realizations of the new hydrogen related
atomic configurations.} \label{illustration.fig}
\end{figure}

In hydrogen embedded platinum junctions a single new configuration
appears, which corresponds to a bridge of a H$_2$ molecule between
the platinum electrodes \cite{Smit2002} [for an illustration see
Fig.~\ref{illustration.fig}(a)] This configuration has a single
channel with perfect transmission as proved by conductance
fluctuation measurements. The interpretation of this result is
supported by theoretical calculations for an arrangement where the
hydrogen molecule lies parallel to the contact axis
\cite{Heurich2003}. Though this configuration is expected to have
a single channel, the perfect transmission is surprising. The high
conductance of H$_2$ can only be understood from the strong
hybridization between the electron states of H$_2$ and the
$d$-band of the electrodes. According to
Ref.~\onlinecite{Heurich2003} the same conductance should be
observed for a hydrogen molecule between palladium electrodes. Our
measurements, however, show that the electrodes cannot be treated
as pure palladium, the dissolved hydrogen has to be taken into
account as well. As shown in Ref.~\onlinecite{Chan1983}, due to
the presence of H atoms in the Pd host the Fermi energy is shifted
upward relative to the $d$-band of Pd. Since in pure Pd the
$d$-band is almost completely filled, the density of states (DOS)
drastically decreases with increasing amount of dissolved
hydrogen. At high H concentration ($\sim 70\%$) the $d$-band
becomes completely closed, and the remaining DOS from the $s$-band
is only $\sim 20\%$ of the original one. On the basis of the
demonstrative model of Ref.~\onlinecite{Heurich2003} the high DOS
at the Fermi energy due to the $d$-electrons is the reason of the
large transmission through the H$_2$ molecule, thus the same
configuration between palladium hydride electrodes should have a
considerably smaller transmission.

A quantitative estimation of the transmission can be given
assuming that the dissolved hydrogen changes only the DOS of the
electrodes. We have performed a calculation with the demonstrative
model of Heurich et al.\ \cite{Heurich2003} applying the DOS of
PdH$_x$ \cite{Chan1983} instead of Pt. Inserting a reasonable
concentration of $x=0.5$ \cite{note2} the model yields a
conductance of $G=0.6\,G_0$.

The above argumentation implies that the configuration with $G\sim
1\,G_0$ is not a hydrogen bridge [Fig.~\ref{illustration.fig}(a)],
like the one in platinum-hydrogen system. This is strongly
supported by our conductance fluctuation measurements as well,
which show that the configuration at $1\,G_0$ has more than one
open conductance channels, whereas the H$_2$ bridge should have a
single channel. On the other hand, the a hydrogen bridge is a
reasonable candidate for explaining the second configuration with
$G\sim 0.5\,G_0$. The estimated conductance value of the H$_2$
bridge is in good agreement with the $\sim 0.5\,G_0$ value, and
the uncertainty of the peak position can be explained by the
changes in the amount of dissolved hydrogen.

We note that in the platinum-hydrogen system the vibrational
spectrum measurements with H$_2$, D$_2$, and HD molecules provided
the conclusion that the new configuration is a molecular and not
an atomic bridge. The same study on Pd contacts is hindered by two
reasons. Due to the unsuppressed conductance fluctuations the
nonlinearity of the $I-V$ curves in atomic-sized contacts is
strongly dominated by quantum interference structures, and the
vibrational spectrum is hardly detectable. Furthermore, the
vibrational modes of a molecular bridge are positioned in the same
energy region as those of the dissolved atoms, thus the separation
of the two features is problematic. Therefore we believe, that the
configuration with $G\sim 0.5\,G_0$ can as well be a molecular as
an atomic hydrogen bridge between palladium-hydride electrodes
[Fig.~\ref{illustration.fig}(a) and (b), respectively].

For the configuration with $G\sim 1\,G_0$ we propose  two atomic
configurations. The first one is a single-atom Pd junction between
Pd-H electrodes [Fig.~\ref{illustration.fig}(c)]. The conductance
of $1.8\,G_0$ for the same configuration between pure electrodes
is shifted close to one quantum unit due to the dissolution of
hydrogen in the electrodes. This reduction of the conductance is
reasonable considering the $30-60\%$ decrease  of the DOS at the
Fermi energy \cite{Chan1983}. This arrangement may have up to $5$
open channels, which agrees with the unsuppressed conductance
fluctuation. A second possibility is the Pd$_2$H$_2$ complex
similarly to that was recently proposed for platinum junctions
\cite{Garcia2004} [Fig.~\ref{illustration.fig}(d)]. In platinum
this arrangement is found to have three open channels with a total
conductance of $1\,G_0$. From these two candidates the monoatomic
Pd contact is supported by the observation that at high hydrogen
concentration the configuration with $G\sim 1\,G_0$ disappears.

In conclusion, we have experimentally investigated the influence
of adsorbed hydrogen on the behavior of atomic-sized palladium
junctions. We have found that the adsorption of hydrogen
completely reshapes the conductance histogram: the original peak
of a monoatomic Pd contact disappears, and two new
hydrogen-related peaks emerge at $G\sim 1\,G_0$ and $G\sim
0.5\,G_0$. Our phonon spectrum measurements have shown that these
configurations are situated between electrodes containing
dissolved hydrogen atoms. The dissolution of hydrogen makes a
crucial difference compared to platinum nanocontacts, where the
hydrogen is found only on the surface. Combining the recent
results on Pt-H$_2$ junctions and the consequences of the
dissolution of hydrogen on the band structure of the Pd
electrodes, we have proposed possible explanations for the new
peaks in the histogram.

This work has been supported by the ``Stichting FOM'' and the
Hungarian research funds OTKA TS040878, T037451. We acknowledge
J.\ Caro for valuable information about his previous work on
palladium-hydride system.


\begin{thebibliography}{19}
\bibitem[{not()}]{note1}
\bibinfo{note}{The typical amount of hydrogen admitted into the sample space
  was $1\,\mu$mole -- $1\,$mmole.}
\end{thebibliography}
\end{document}